\documentstyle[preprint,eqsecnum,aps]{revtex}

\newcommand{\note}[1]{}
\renewcommand{\note}[1]
               {\marginpar{Note}
               ({\bf Note: }{\em {#1})\ }}
\newcommand{\reply}[1]{}
\renewcommand{\reply}[1]
                {({\bf Reply: }{\tt {#1})\ }}
\newcommand{\tauint}{\tau_{int}}
\newcommand{\tauexp}{\tau_{exp}}
\newcommand{\la}{\langle}

\newcommand{\ra}{\rangle}

\newcommand{\e}{{\rm e}}

\newcommand{\beq}{\begin{equation}}
\newcommand{\eeq}{\end{equation}}
\newcommand{\beqa}{\begin{eqnarray}}
\newcommand{\eeqa}{\end{eqnarray}}

\newcommand{\ssc}[1]{Subsection~\ref{#1}}

\newcommand{\Eq}[1]{(\ref{#1})}

\newcommand{\Fig}[1]{Fig.~\ref{#1}}

\newcommand{\Table}[1]{Table~\ref{#1}}
\newcommand{\tab}[1]{Table~\ref{#1}}

\newcommand{\th}{{\rm tanh}}

\newcommand{\algP}{$P$}
\newcommand{\algL}{$L$}
\newcommand{\algLex}{$L_{\rm ex}$}

\newcommand{\nup}{n^{\uparrow}}
\newcommand{\ndn}{n^{\downarrow}}

\newcommand{\phiup}{\phi^{\uparrow}}
\newcommand{\phidn}{\phi^{\downarrow}}
\newcommand{\psiup}{\psi^{\uparrow}}
\newcommand{\psidn}{\psi^{\downarrow}}

\newcommand{\PL}{$PL$}
\newcommand{\PLex}{$PL_{\rm ex}$}
\newcommand{\LLex}{$LL_{\rm ex}$}
\newcommand{\PLLex}{$PLL_{\rm ex}$}

\begin{document}

\title{
Loop algorithms for quantum simulations of \\
fermion models on lattices
}

\author{N.\ Kawashima and J.E.\ Gubernatis}
\address{
Center for Nonlinear Studies and Theoretical Division\\
Los Alamos National Laboratory, Los Alamos, NM 87545
}

\author{H.G.\ Evertz}
\address{
Center for Simulational Physics, Department of Physics\\
University of Georgia, Athens, GA 30602 \\
{\em and}\\
Supercomputer Computations Research Institute\\
Florida State University, Tallahassee, FL 32306 \\
}

\maketitle

\newpage

\begin{abstract}
Two cluster algorithms, based on constructing and flipping
loops, are presented for worldline quantum Monte Carlo simulations of
fermions and are tested on the one-dimensional repulsive Hubbard
model.  We call these algorithms the loop-flip and loop-exchange
algorithms. For these two algorithms and the standard worldline
algorithm, we calculated the autocorrelation times for various
physical quantities and found that the ordinary worldline algorithm,
which uses only local moves, suffers from very long correlation times
that makes not only the estimate of the error difficult but also the
estimate of the average values themselves difficult.  These
difficulties are especially severe in the low-temperature, large-$U$
regime. In contrast, we find that new algorithms, when used alone
or in combinations with themselves and the standard algorithm, can
have significantly smaller autocorrelation times, in some cases being
smaller by three orders of magnitude.  The new algorithms, which use
non-local moves, are discussed from the point of view of a general
prescription for developing cluster algorithms.  The loop-flip
algorithm is also shown to be ergodic and to belong to the grand
canonical ensemble.  Extensions to other models and higher dimensions
is briefly discussed.

\begin{center}
(To appear in Phys. Rev. B)
\end{center}
\end{abstract}


\newpage

%%%%%%%%%%%%%%%%%%%%%%%%%%%%%%%%%%%%%%%%%%%%%%%%%%%%%%%%%%%%%%%%%%
%                                                                %
%                   INTRODUCTION                                 %
%                                                                %
%%%%%%%%%%%%%%%%%%%%%%%%%%%%%%%%%%%%%%%%%%%%%%%%%%%%%%%%%%%%%%%%%%
\section{Introduction}
\label{sec:Introduction}

The worldline quantum Monte Carlo method is frequently used by
condensed matter and field theorists to simulate lattice models of
systems of interacting fermions, bosons, and quantum spins
\cite{suzuki87}.  This method has become a textbook example
\cite{negle88} of a quantum Monte Carlo method as one of its virtues
is its simplicity. Another virtue is its production of worldline
patterns that often pictorially represent the imaginary-time quantum
dynamics of the model.  Several difficulties with the method are also
well known.

The most notable difficulty \cite{hirsch81}, which the worldline
method shares with almost all other quantum Monte Carlo methods, is a
sign problem which is manifested by Monte Carlo transition
probabilities becoming negative. Typically, this difficulty renders
the method ineffective. A difficulty \cite{hirsch81} more unique to
the worldline method is the lack of ergodicity as in practice the
winding number of the worldlines is conserved and thereby the sampling
of phase space is restricted.  The winding number conservation
corresponds to the conservation of fermion number and thus places the
method in the canonical ensemble.

A less appreciated and infrequently studied difficulty is the
worldline method's very long auto-correlation time between
measurements of physical quantities.  These long times make error
estimation for these quantities difficult and can cause long computer
runs. The main purpose of this paper is to illustrate the
extraordinary lengths these times can take and to present two new ways
of implementing the worldline method that in many cases reduce these
times by several orders of magnitude.  The new methods are a
significant improvement in efficiency.  One method is also naturally
ergodic and grand canonical as winding number is not conserved.

With improved efficiency, the new methods generate significant
reductions in the variance of the calculated results for a fixed
amount of computing time.  Since the sign problem is usually
accompanied by a dramatic increase in the variance of the measured
quantities, the extensions potentially mitigate the sign problem.
They do not, however, address it directly. Oddly, some combinations of
the new methods permit a sign problem to occur in simulations for
which the standard implementations have no sign problem.  We found
that this was a minor problem, occurring infrequently only at high
temperature in very small systems and decreasing as the lattice size
was increased.  Constructing still other methods that conserve winding
number is possible; however, allowing the winding number to change
often appears to be a key ingredient for improved performance.

We will present our new algorithms by discussing their application to
worldline simulations of the one-dimensional repulsive Hubbard model.
The Hubbard model is one of the simplest models of interacting
electrons, and for it we will study the auto-correlation time among
measurements of the different physical quantities germane to its
interesting physics.  In Section II, we define the model and give a
brief description of the computation of its thermodynamic properties
from a path-integral.  Here, we will also define and discuss how we
measured the auto-correlation times.  To establish notation and make
subsequent discussion reasonably self-contained, we also present the
particular implementation of the worldline method used in our studies.
This implementation focuses on Monte Carlo moves that change the state
of plaquettes defined on the space-time lattice on which the
worldlines exist.

In Section III, we present our two new worldline methods, which we
call the loop-flip and loop-exchange methods. The loop-flip method is
an extension of the work of Evertz et al.\ \cite{evertz93} who
developed a loop-flip algorithm to reduce critical slowing down in
6-vertex model simulations.  In that work, they observed that
worldline simulations of quantum spin systems have similarities to
simulations of the 6-vertex model and suggested the potential utility
of their algorithm for simulating such systems.  We will discuss the
extensions of the procedure necessary for simulating fermion systems.
We also present our second method, the loop-exchange algorithm.  For
fermions systems, we found it very useful to exchange portions of up
and down spin worldlines to accelerate the sampling of phase space.
This exchange was designed to overcome the difficulty of moving up and
down spin worldlines across one another because of the Coulomb
repulsion between the fermions that exists in the Hubbard model.  Our
two new algorithms, in sharp contrast with the standard
implementations of the worldline method, use non-local (global)
updating moves.  These moves generalize the cluster algorithms
recently developed to reduce long auto-correlation times accompanying
simulations of critical phenomena in classical spin systems
\cite{kandel91}.  Here, we are not concerned with critical phenomena,
but rather we are reducing the inherently long auto-correlation times
that occur in the worldline method even when the physical system is
far removed from any known phase transition.

We investigate in Section IV the effectiveness of various combinations
of the algorithms for computing several different properties of the
repulaive Hubbard model.  For a standard version of the worldline
method, which we call the plaquette-flip algorithm, the
auto-correlation time was too long to measure in most cases, even with
the use of relatively long computer runs.  In these cases, reliable
error estimation is very difficult.  Our new algorithms did not suffer
from this problem.  Finally, in Section V, we summarize our findings
and discuss important issues awaiting further investigation,
particularly for doing models other than the Hubbard model and for
studying models in higher dimensions.

%%%%%%%%%%%%%%%%%%%%%%%%%%%%%%%%%%%%%%%%%%%%%%%%%%%%%%%%%%%%%%%%%%
%                                                                %
%               Background and Definitions                       %
%                                                                %
%%%%%%%%%%%%%%%%%%%%%%%%%%%%%%%%%%%%%%%%%%%%%%%%%%%%%%%%%%%%%%%%%%
\section{Background and Definitions}
\label{sec:Background}

\subsection{Hubbard Model}
\label{subsec:Hubbard}
The one-dimensional Hubbard Hamiltonian is
\begin{equation}
H \equiv \sum_{i} H_{i,i+1}
               =\sum_{i}[T_{i,i+1} + \frac{1}{2}(V_i+V_{i+1})]
\label{eq:Hubbard}
\end{equation}
where
\begin{equation}
T_{i,i+1} = -t\sum_\sigma(c_{i,\sigma}^{\dagger}c_{i+1,\sigma}+
                      c_{i+1,\sigma}^{\dagger}c_{i,\sigma})
\label{eq:Kinetic}
\end{equation}
and
\begin{equation}
V_i= \sum_\sigma \Bigl[\frac{1}{2} U (n_{i,\sigma}-{1\over 2})
                    (n_{i,-\sigma}-{1\over 2})
    - \mu (n_{i,\sigma}-{1 \over 2}) \Bigr]
\label{eq:Potential}
\end{equation}
Here, $c_{i,\sigma}^{\dagger}$ and $c_{i,\sigma}$ are the creation and
destruction operators for a fermion at site $i$ with spin $\sigma$ (up
or down) and $n_{i,\sigma} = c_{i,\sigma}^{\dagger}c_{i,\sigma}$ is
the fermion number operator at site $i$ for spin $\sigma$. The first
term in (\ref{eq:Hubbard}) is the kinetic energy of the electrons and
describes their hopping, without spin flip, from site to site. The
second term is the potential energy (Coulomb interaction) that exists
only if two electrons occupy the same site and includes the chemical
potential $\mu$ term for convenience. We took $U>0$. The model is
defined in such a way that when $\mu$ equals zero, the model exhibits
particle-hole symmetry.  With this symmetry, there are an equal number
of up and down spin electrons and their sum equals the number of
lattice sites $N$ (the half-filled case). We assume periodic boundary
conditions.

At half-filling, the system is an insulator, with anti-ferromagnetic
spin fluctuations dominating charge-density wave fluctuations. At
other fillings, it is metallic. Increasing the parameter $U$ supresses
the charge fluctuations and enhances the spin fluctuations. The
following correlation functions are useful descriptors of these two
types of fluctuations \cite{hirsch83},
\begin{equation}
 T\chi_{\pm}(q) = \frac{1}{\beta N}\int_0^\beta d\tau \sum_{j,k} e^{i q\cdot k}
         \langle (n_{j+k,\uparrow}(\tau) \pm n_{j+k,\downarrow}(\tau))
                 (n_{j,\uparrow}(0) \pm n_{j,\downarrow}(0) \rangle
\label{eq:Chi}
\end{equation}
They measure static spin-density wave (SDW) and charge-density wave
(CDW) correlations.  Simpler measurable quantities include the energy
$E=\langle H \rangle$ and the average electron occupancy per site,
\begin{equation}
  n = \frac{1}{\beta N}\int_0^\beta d\tau
     \sum_j \langle(n_{j\uparrow}(\tau) + n_{j\downarrow}(\tau))\rangle
\label{eq:N}
\end{equation}
As we discuss below, we compute these quantities as functions of
imaginary-time $\tau$ and their definitions reflect an average over
this parameter.

In (\ref{eq:Chi}) and (\ref{eq:N}), the symbol $\langle \cdots
\rangle$ denotes the finite-temperature expectation value of some
physical observable and is defined by
\begin{equation}
  \langle A \rangle = {\rm Tr\,} A e^{-\beta H} / Z
\label{eq:Expectation}
\end{equation}
where $\beta=1/kT$ is the inverse temperature, $Z = {\rm Tr\,}
e^{-\beta H}$ is the partition function of the system, and Tr denotes
the trace operation.  The basic idea of the worldline method is to
express the trace operation as a path-integral in imaginary-time and
then use Monte~Carlo techniques to evaluate the resulting
multi-dimensional functional~integral numerically.

\subsection{Path-Integrals}
\label{subsec:Path}

To develop the path-integral, we rewrite the partition function of the
system by dividing the imaginary-time interval
$0\!\le\!\tau\!<\!\beta$ into $L$ slices of width $\tau=\beta/L$ and
inserting complete sets of states $|s\rangle$ at each time-slice:
\begin{equation}
  Z = \sum_{\{s_i\}}
        \langle s_1|e^{-\tau H}|s_L\rangle
        \cdots
        \langle s_3|e^{-\tau H}|s_2\rangle
        \langle s_2|e^{-\tau H}|s_1\rangle
\label{eq:partition}
\end{equation}
where the sum is over the set $\{s_i\}=\{s_1,s_2,\dots,s_L\}$ of all
states $s_i$. The subscript on the identifier $s$ of the different
members of the complete set marks the time-slice at which the states
were inserted.  The properties of the trace require the propagation in
imaginary-time to be periodic.

In order to simplify the evaluation of the matrix elements in
(\ref{eq:partition}), we rewrite the full Hamiltonian as the sum of
two easily diagonalizable, but not necessarily commuting, pieces, $H =
H_1+ H_2$.
Using one form of the Suzuki-Trotter approximation \cite{suzuki76}, we may
then approximate the imaginary-time evolution operator $e^{-\tau H}$
by
\begin{equation}
 e^{-\tau H} \approx e^{-\tau H_2}e^{-\tau H_1}
\label{eq:trotter}
\end{equation}
When this approximation is used in the expression for the partition
function (\ref{eq:partition}) and additional intermediate states are
inserted, the resulting expression for the partition function is
\begin{eqnarray}
    Z &=& \sum_{\{s_i\}}
          \langle s_1|e^{-\tau H_2}|s_{2L}\rangle
          \cdots
          \langle s_4|e^{-\tau H_1}|s_3\rangle \nonumber \\
      & & \quad \quad \times
          \langle s_3|e^{-\tau H_2}|s_2\rangle
          \langle s_2|e^{-\tau H_1}|s_1\rangle
\label{eq:Partition}
\end{eqnarray}
In Section \ref{subsec:Worldlines}, we will give specific choices of $H_i$ and
$|s\rangle$ for the Hubbard model.  Here, we are concerned only
with the general framework of the method.

We now see that the partition function (\ref{eq:Partition}) may be
re-expressed as a functional integral over all possible configurations
${\cal C}\equiv\{s_i\}$.  To evaluate the
path integral, Monte~Carlo importance sampling is used.  This method
generates a sequence of configurations ${\cal C}_i$ (see below),
which, in the infinite-sequence limit, conforms to the probability
distribution
\begin{eqnarray}
  P({\cal C}) &=& \langle s_1|e^{-\tau H_2}|s_{2L}\rangle \cdots
          \langle s_3|e^{-\tau H_2}|s_2\rangle \langle s_2|e^{-\tau
H_1}|s_1\rangle /Z \nonumber \\
              &\equiv& W({\cal C})/Z
\label{eq:Weight}
\end{eqnarray}
Such a sequence may be generated by beginning with an arbitrary
(allowed) worldline configuration ${\cal C}_0$ and then considering
some modified configuration ${\cal C}'_0$.  One then computes the
probabilities for these configurations, $P({\cal C}_0)$ and $P({\cal
C}'_0)$, and accepts the modified configuration with probability $R$
for which we choose
\begin{equation}
 R = {{P({\cal C}'_0)} \over {P({\cal C}_0) +P({\cal C}'_0)}}
   = {{W({\cal C}'_0)} \over {W({\cal C}_0) + W({\cal C}'_0)}}
\label{eq:HeatBath}
\end{equation}
If the modification is accepted, the new configuration will be ${\cal
C}_1={\cal C}'_0$; otherwise the old configurations will be retained:
${\cal C}_1={\cal C}_0$.  The procedure is then iterated to produce
${\cal C}_2$, ${\cal C}_3$, etc.

\subsection{Worldlines}
\label{subsec:Worldlines}

For fermion lattice models, the standard choice for the complete set of
states $|s\rangle$ is the occupation number basis formed from the
state $|n_{1,\uparrow},n_{2,\uparrow},\dots,n_{n,\uparrow}\rangle
|n_{1,\downarrow},n_{2,\downarrow},\dots,n_{N,\downarrow}\rangle$ by
allowing each lattice site to assume all possible occupancies of up
and down electrons \cite{hirsch81}.  The splitting of the Hamiltonian
$H = H_1 + H_2$ is usually done by having $H_1$ and $H_2$ refer to the
odd and even lattice sites \cite{hirsch81}. With these choices, the
product of the matrix elements in the expression (\ref{eq:Partition})
for the partition function factorizes into products of matrix elements
defined on shaded squares (plaquettes) of up and down spin
checkerboards.  The summation over all states becomes a summation over
all combinations of occupancies $n_{i,l}^\sigma$ for each spin.  Here,
$i$ labels the spatial (real space) position and $l$, the temporal
(imaginary-time) position.  A shaded plaquette for a given spin is one
with the four sites $(i,l)$, $(i+1,l)$, $(i,l+1)$, and $(i+1,l+1)$
where $i$ and $l$ are either both even or both odd.  A worldline for a
given spin is a continuous line constructed from straight line
segments connecting occupied sites.  A one-to-one correspondence
between fermions and worldlines can be made if two parallel vertical
(temporal) lines are assigned to a shaded plaquette with four occupied
sites. These concepts are illustrated in Fig.~\ref{fg:1} which shows
the checkerboard and a typical worldline configuration for one spin
component of a system of 4 electrons on an 8 site lattice.

Because of the checkerboarding, we can write the partition function as
\begin{equation}
  Z =  \sum_{\{n^{\sigma}_{i,l}\}}
  {\rm sgn}(\{n^{\sigma}_{i,l}\})
  W(\{n^{\sigma}_{i,l}\})
\label{eq:Z}
\end{equation}
\begin{eqnarray}
  W(\{n^{\sigma}_{i,l}\}) =
  \prod_{l=1}^{L}\prod_{i=1}^{N/2}
  &w& \left(
    \begin{array}{ll|ll}
      \nup_{2i-1,2l  } & \nup_{2i  ,2l  } &
      \ndn_{2i-1,2l  } & \ndn_{2i  ,2l  } \\
      \nup_{2i-1,2l-1} & \nup_{2i  ,2l-1} &
      \ndn_{2i-1,2l-1} & \ndn_{2i  ,2l-1} \\
    \end{array}
    \right)  \nonumber \\
  & & \nonumber \\
  \times
  &w& \left(
    \begin{array}{ll|ll}
      \nup_{2i  ,2l+1} & \nup_{2i+1,2l+1} &
      \ndn_{2i  ,2l+1} & \ndn_{2i+1,2l+1} \\
      \nup_{2i  ,2l  } & \nup_{2i+1,2l  } &
      \ndn_{2i  ,2l  } & \ndn_{2i+1,2l  } \\
    \end{array}
    \right)
\label{Bfactor}
\end{eqnarray}
The arguments of the functions on the right-hand side are written in a
form that emphasizes the plaquette structure for both up and down
spins.  Clearly, the weight (\ref{Bfactor}) reduces to a product of
weights for only the shaded plaquettes.
If we further choose
\begin{equation}
  e^{-\tau H_{i,i+1}} \approx e^{-\tau(V_i + V_{i+1})/4}
  e^{-\tau T_{i,i+1}} e^{-\tau(V_i + V_{i+1})/4}
\end{equation}
then we can express the local weights as
\begin{eqnarray}
  w \left(
    \begin{array}{ll|ll}
      \nup_1 & \nup_2 &
      \ndn_1 & \ndn_2 \\
      \nup_3 & \nup_4 &
      \ndn_3 & \ndn_4 \\
    \end{array}
    \right)
    & \equiv &
    w_T \left(
    \begin{array}{ll}
      \nup_1 & \nup_2 \\
      \nup_3 & \nup_4 \\
    \end{array}
    \right)
    w_T \left(
    \begin{array}{ll}
      \ndn_1 & \ndn_2 \\
      \ndn_3 & \ndn_4 \\
    \end{array}
    \right) %\nonumber \\
    w_V \left(
    \begin{array}{ll|ll}
      \nup_1 & \nup_2 &
      \ndn_1 & \ndn_2 \\
      \nup_3 & \nup_4 &
      \ndn_3 & \ndn_4 \\
    \end{array}
    \right)
\label{eightbodyw}
\end{eqnarray}
with
\begin{eqnarray}
  w_T \left(
    \begin{array}{ll}
      n_1 & n_2 \\
      n_3 & n_4 \\
    \end{array}
    \right)
  & \equiv &
      \delta_{n_1,n_2}\delta_{n_3,n_4}\delta_{n_1,n_3}
         + \cosh(\tau t)
      \delta_{n_1,1-n_2}\delta_{n_3,1-n_4}\delta_{n_1,n_3}
                                 \nonumber \\
   & & \quad + \sinh(\tau t)
      \delta_{n_1,1-n_2}\delta_{n_3,1-n_4}\delta_{n_1,1-n_3}
\label{eq:wt}
\end{eqnarray}
and
\begin{equation}
    w_V \left(
    \begin{array}{ll|ll}
      \nup_1 & \nup_2 &
      \ndn_1 & \ndn_2 \\
      \nup_3 & \nup_4 &
      \ndn_3 & \ndn_4 \\
    \end{array}
    \right)
    = \prod_{i=1}^4 w_v(\nup_i,\ndn_i)
\end{equation}
where
\beq
  w_v(n,n') \equiv \e^{-\frac{\tau}{4} [U (n-\frac1{2})(n'-\frac1{2})
                   -\mu (n+n'-1)]}
\eeq
In (\ref{eq:wt}), we see the kinetic energy's contribution to the
weight is non-vanishing only for certain sets of occupancies.  In
fact, out of the $2^4=16$ possible plaquettes for either an up or a
down spin, only 6 plaquettes satisfy local fermion conservation and
these are shown in Fig.~\ref{fg:2}.

The manifold on which the worldlines are defined is a two-dimensional
torus since we have a periodic boundary condition for both spatial and
temporal directions.  Accordingly, on this torus, we can define
spatial and temporal winding numbers for any closed loop, for example,
a worldline.  To be more specific, to define a spatial winding number
for a given loop, let us first suppose we have a spatial cut, i.e., a
vertical line in Fig.1 which starts somewhere at the bottom and ends
at the top of the checkerboard.  (The location of the cut does not
matter since it does not affect the definition.) Then, if we trace the
entire path of the loop and it passes a spatial cut in one direction
$m$ times and in the opposite direction $n$ times, then the spatial
winding number of this loop is defined to be $m-n$.  The total spatial
winding number of worldlines is defined as the sum of the spatial
winding numbers of all the worldlines.  A temporal winding number of a
given loop and the total temporal winding number of worldlines are
defined in the same fashion but with the temporal cut being a
horizontal line in the checkerboard.  In what follows, when we say {\it
spatial or temporal winding number}, we will mean the total spatial or
temporal winding number of the worldlines.

The sign in (\ref{eq:Z}) is defined in terms of the worldlines
\beq
  {\rm sgn}(\{n^{\sigma}_{i,l}\})
  \equiv
  (-1)^{\sum_l (Z_l-1)}
\label{eq:Sign}
\eeq
where $Z_l$ is the temporal winding number of the $l$-th worldline and
the sum is taken over all the worldlines. Its origin is the fermion
anti-commutation. For the one-dimensional Hubbard model, the standard
worldline algorithms have no sign problem as $Z_l=1$ and is conserved.
For the new algorithms we will be proposing, this will not be the
case.  Whenever a sign problem occurred in our simulations, we treated
it in the ordinary way \cite{blankenbecler81}; that is, when we have
configurations of negative sign, we computed thermodynamic averages by
\beq
   \la A \ra =
   \sum_{{\cal C}}{\rm sgn}({\cal C}) A({\cal C}) /
   \sum_{{\cal C}}{\rm sgn}({\cal C})
\eeq
By using this formula, we do the Monte Carlo simulation
with the following weight
\beq
  Z'  =  \sum_{\{n^{\sigma}_{i,l}\}} W(\{n^{\sigma}_{i,l}\}).
\eeq
that ignores the sign factor.

\subsection{The plaquette-flip algorithm (Algorithm \algP)}
\label{subsec:P}

The worldline method can be implemented in several different ways.  We
will now highlight the main points of an implementation which focuses
on Monte Carlo moves that change the local fermion occupancies of
plaquettes defined on a checkerboarding of imaginary-time and space.
This implementation leads to efficient algorithms not only on
serial computers but also on vector and parallel machines
\cite{okabe86,marcu87,somsky92}.

To generate different sequences of worldlines, one uses the
observation \cite{hirsch81} that only one basic movement of a
worldline segment can occur: If an unshaded plaquette has a worldline
segment on one vertical edge and none on the other, the allowed
movement is vacating the one edge and occupying the other by moving
the worldline segment across the unshaded plaquette.  This movement is
illustrated in Fig.~\ref{fg:3}.  Algorithm \algP\ accomplishes this by
visiting in sequence each unshaded plaquette for both the up and down
spin checkerboards and attempting a local move that flips all four
corners of the unshaded plaquette.  {\it Flipping a corner} means
replacing the value of the variable $n_{i,l}$ by $1-n_{i,l}$.  The
configuration weights $W({\cal C})$ and $W({\cal C}')$ are determined
by computing the product of the weights (\ref{eightbodyw}) for the
four neighboring shaded plaquettes for both the old and flipped
configuration.  These weights are then used in (\ref{eq:HeatBath}) to
determine whether the flip is accepted or not.  An important point is
that we need to know only local states (the occupancies of the shaded
plaquettes surrounding the unshaded one) to calculate this probability
rather than the state of the whole system.  This is the case because
the attempted flip changes the configuration only locally since the
Boltzmann weight (\ref{eightbodyw}) is factorized into a product of
local weights .

In algorithm \algP, the winding number of any worldline in any
direction, temporal or spatial, is conserved.  We found, however, that
this broken ergodicity, as expected \cite{hirsch81}, has only a small
effect on the simulation, especially for larger lattices.

\subsection{Auto-correlation times}
\label{subsec:Times}

The basic property of the Monte Carlo method is to replace the
thermodynamic average (\ref{eq:Expectation}) by the average of $A(t)$
over $M$ Monte Carlo steps:
\begin{equation}
  \langle A \rangle \approx {\bar A} \equiv {1\over M}\sum_{t=1}^M A(t)
\end{equation}
where $A(t)$ is the value of a physical quantity $A$ (energy, SDW,
etc.) at Monte Carlo step $t$.  If $M$ is large enough and the
$A(t)$'s are {\it statistically independent} estimates of $A$, then
the error estimate for $A$ would be $\sigma /\sqrt{M}$ where
\begin{equation}
  \sigma^2 = {1 \over{M-1}}\sum_{t=1}^M (A(t) - \bar A)^2
\label{eq:Sigma}
\end{equation}

To measure the degree of statistical correlations, we calculated two
kinds of quantities.  The first one is an integrated autocorrelation
time defined by
\begin{equation} \label{eq:discTauInt}
  \tau_{\rm int}^A \equiv -\frac{1}{2}+\sum_{t=0}^\infty \Gamma_A(t).
\end{equation}
where
\beq
  \Gamma_A(t) \equiv
  \frac{\la A(t_0 + t) A(t_0) \ra - \la A(t_0) \ra \la A(t_0) \ra}
  {\la A(t_0    ) A(t_0) \ra - \la A(t_0) \ra \la A(t_0) \ra}.
\eeq
In the actual simulation, $\Gamma_A(t)$ is approximated by
\begin{equation} \label{eq:Gamma}
\Gamma_A(t) \approx
%
%  {{{1 \over{M-t}}\sum_{i=1}^{M-t} (A(i+t)A(i) - \bar A\bar A)}
%
  {{{1 \over{M-t}}\sum_{i=1}^{M-t} A(i+t)A(i) -
    {1 \over{(M-t)^2}} \left[\sum_{i=  1}^{M-t} A(i)\right]
                       \left[\sum_{i=t+1}^{M  } A(i)\right] }
                  \over
  {{1 \over M}\sum_{i=1}^M [A(i) A(i) - \bar A \bar A]}}
\end{equation}
The variance $\sigma_{int}^2$ of data correlated with the time
$\tau_{int}^A$ is related to the variance $\sigma^2$ computed from
(\ref{eq:Sigma}) in the absence of correlations by \cite{binder88}
\begin{equation}
 \sigma_{int}^2 \approx 2\tau_{int}^A\sigma^2
\label{eq:sigmaint}
\end{equation}
Thus, in the presence of correlations, $2\tau_{int}^A$ more
simulation steps are needed in order to achieve the same variance of a
measured quantity.  Accordingly, error estimates computed when
correlations are unknowingly present are always smaller than the
actual error.

Typically, the auto-correlation function is not a simple exponential
function in time.  This fact sometimes makes estimating $\tau_{int}^A$
by (\ref{eq:discTauInt})
difficult.  Another estimate of $\tau_{int}^A$ is obtained by
examining what it takes to
produce statistically independent measurements.  We grouped the $M$
measurements into $n
$ bins of length $l= M/n$, and for a sequence of
bin lengths (chosen to be 2, 4,8, $\dots$),
 we computed the bin averages $\bar A_b(l)$ of the $A(t)$
\beq
  \bar A_b(l) \equiv \frac1{l}\sum_{t=(b-1)l+1}^{bl} A(t).
\eeq
and the variance of these averages
\begin{equation}
  \sigma(l)^2 \equiv {1 \over{n - 1}}\sum_{b=1}^n
                                        (\bar A_b(l) - \bar A)^2
\end{equation}
This variance should become inversely proportional to $l$ as the bin
size $l$ becomes large enough so that the $\bar A_b(l)$ as a function
of $b$ become statistically independent \cite{allen87}.  When
statistical independence is approached, the quantity
\begin{equation}
  \tauint^A(l) = { {l \sigma(l)^2} \over 2 \sigma^2}
\label{eq:TauBin}
\end{equation}
where $\sigma^2$ is computed from (\ref{eq:Sigma}), approaches a
constant as a function of $l$ \cite{allen87}.  This constant is our
estimate for $\tauint^A$.  The error estimate becomes
$\sigma(l^*)/\sqrt{n}$ where $l^*$ is the value of $l$ at which the
constant is reached. We note that with the use of (\ref{eq:TauBin}),
we can rewrite this error estimate as $\sigma\sqrt{2\tauint/M}$ and
obtain the expected result (\ref{eq:sigmaint}). We also note that we
have defined $\tauint^A(l)$ so that $\tauint^A(1)=\frac{1}{2}$.

Generally, $\tauint^A(l)$ is an increasing function of $l$. This
increase, however, ceases when $l$ becomes much larger than the
auto-correlation time $\tauint^A$. As we discuss below, we found cases
where, in spite of a very long computer run, $\tauint^A(l)$ keeps
increasing as function of $l$. In such cases, we estimated a lower
bound for $\tauint^A$ by selecting the value $\tauint^A(l)$ for the
largest available $l$.

A different measure of auto-correlation time we used is the exponential
auto-correlation time $\tauexp$ which by definition
is the largest autocorrelation time associated with a given algorithm.
We determined $\tauexp$ from the ${t
\rightarrow \infty}$ asymptotic behavior of $\Gamma_A(t)$,
\begin{equation}
  \Gamma_A(t) \sim \gamma_A \exp(-t/\tauexp),
\label{TauExp}
\end{equation}
by fitting a straight line to $\log \Gamma_A(t)$. We remark that in
practice, the most stable estimates of $\tauint^A$ were obtained by
summing $\Gamma_A(t)$ in (\ref{eq:discTauInt}) only up to some finite
$t=W$, with $W$ of the order of $\tauexp$, and computing the
contribution of $t> W$ from (\ref{TauExp}).

The auto-correlation time $\tauexp$ corresponds to the second largest
eigenvalue of the transition probability matrix $W({\cal
C}\rightarrow{\cal C}')$ for changing configuration ${\cal C}$ to
${\cal C}'$.  (The largest eigenvalue is one, with the Boltzmann
distribution as the eigenvector \cite{hammersley64}.)
This correlation time depends on
the algorithm. The constant $\gamma_A$ in (\ref{TauExp}) depends on
the observable $A$ (as well as on lattice size, temperature, etc.) and
can be very small or zero for some observables.  In general, $\tauexp$
is more difficult to estimate precisely than $\tauint^A$. In order to
determine accurately $\tauexp$, we must therefore use a suitable range
of observables to establish consistency.  In some cases, we could not
obtain reliable estimates of $\tauexp$ as we will see below. Another
remark about $\tauexp$ is that the longest-lived physical mode associated
with it can be a mode whose contributions to the quantities of
interest are negligible. We will discuss this point more fully later.

%%%%%%%%%%%%%%%%%%%%%%%%%%%%%%%%%%%%%%%%%%%%%%%%%%%%%%%%%%%%%%%%%%
%                                                                %
%                    A L G O R I T H M S                         %
%                                                                %
%%%%%%%%%%%%%%%%%%%%%%%%%%%%%%%%%%%%%%%%%%%%%%%%%%%%%%%%%%%%%%%%%%
\section{New Algorithms}
\label{sec:Algorithms}

%%%%%%%%%%%%%%%%%%%%%%%%%%%%%%%%%%%%%%%%%%%%%%%%%%%%%%%%%%%%%%%%%%
%%%%%%%%   L o o p   F L I P   A L G O R I T H M   %%%%%%%%%%%%%%%
%%%%%%%%%%%%%%%%%%%%%%%%%%%%%%%%%%%%%%%%%%%%%%%%%%%%%%%%%%%%%%%%%%
\subsection{The loop-flip algorithm (Algorithm \algL)}
\label{subsec:L}

We now discuss the loop-flip algorithm (algorithm \algL).  In contrast
to algorithm \algP, which makes only local changes in configuration
space based on local decisions, the loop-flip algorithm makes
non-local moves based on local decisions that are linked in a specific
manner.  The manner chosen is related to the loop-flip algorithm
recently proposed by Evertz et al.\ \cite{evertz93} to reduce critical
slowing down in simulations of the 6-vertex model.  The formal
connection of the six-vertex model with the fermion (and quantum spin)
problem is through the shaded plaquettes for each spin assuming only 6
out of the 16 possible configurations. The main conceptual connection
is through the recognition that the worldline and the six-vertex
configurations \cite{baxter82} can be parameterized by closed loops as
a consequence of a zero-divergence property of the models. In the fermion
case, this property is directly connected with the local conservation
of fermions.  From the loop point of view, the difference between two
different worldline configurations must simply be one or more closed
loops as the difference must also satisfy the zero-divergence
condition.  It is such differences that the loop-flip algorithm
constructs.

The core of our loop-flip algorithm is the ``massless'' case of the
6-vertex algorithm proposed by Evertz et al.\ \cite{evertz} which uses
only ``break-up'' operations.  For each spin (up or down), the 6
allowed plaquettes are mapped onto 3 new plaquettes that have lines
drawn on them.  Since each lattice site belongs to two shaded
plaquettes, these lines (loop segments) are drawn so that each site is
touched by one line, i.e., we have zero divergence.  When these lines
are connected, loops form, some of which are very long and wind one or
more times around the temporal and spatial directions.  For each spin,
flipping the loop corresponds to changing electrons on the loop into
holes and vice versa.  In the process, the number of electrons can
change.  This change occurs when the net temporal winding number
associated with the system changes. If the temporal winding number
changes, a ``sign problem'' may occur (\ref{eq:Sign}).

The existence of just 6 allowable plaquettes for each spin does not
mean the Hubbard model is equivalent to a combination of 6-vertex
models.  Several special considerations exist. In (\ref{Bfactor}), the
presence of the sign term is one.  As we mentioned above, our use
of the loop-flip algorithm can generate negative weights; fortunately,
we found them for only uninteresting cases and then very rarely.  A
second consideration is the $U$-term and $\mu$-term in the Hubbard
model. We take these terms into account by observing that in their
absence the weight (\ref{eightbodyw}) factors into an up and down
term, each of which maps onto a 6-vertex model.  We apply our
loop-flip algorithm to each of these pieces and then include the $w_V$
contribution by modifying the acceptance probability so that the
detailed balance condition is satisfied.  This modification is
straightforward; our main task is specifying the loop-flip algorithm.

To construct our algorithm, we first imagine that we stack the up
and down spin checkerboards slightly above one another and focus our
attention on blocks whose upper and lower faces are the up and down
spin shaded plaquettes. (We have 36 allowed local configurations for
each block.)  Next, we define the variable $\phi_b$ which specifies
the state of a block $b$ on which the eight-body local interaction in
\Eq{eightbodyw} is defined.  For this variable, we take a pair of
symbols, $\phi_b \equiv (\phiup_b, \phidn_b)$, each of which has one
of the six allowed values of the plaquette. We symbolize these six
values by $1$,$\bar 1$, $2$, $\bar 2$, $3$, and $\bar 3$ and define them in
\Fig{fg:2}.  We then rewrite the Boltzmann weight (\ref{eightbodyw}) in
terms of local variables defined on the shaded blocks as follows:
\beqa
  W({\cal C}) & \equiv & \prod_b w(\phi_b), \\
  w(\phi_b) & = & u(\phidn_b) u(\phiup_b) v(\phidn_b,\phiup_b).
\label{localw}
\eeqa

Our task is to generate a Markov process, in which different
configurations are visited with a frequency proportional to the weight
(\ref{localw}).  This can be done by cycling a Monte Carlo updating
procedure that consists of two steps \cite{kandel91}: The first step
is the stochastic assignment of a label to every shaded block, and the
second step is a Monte Carlo move using modified weights associated
with these labels.

In the first step, we assign a label $\psi_b$ to each shaded block
with a probability $p(\psi_b|\phi_b)$. Here, the label $\psi_b$ is a
pair of integers, $(\psidn_b,\psiup_b)$, each of which may have a
value 1, 2 or 3.  These labels will be related to decompositions of a
plaquette into line segments.  (See Fig.~\ref{fg:L}.) The
probabilities are to be determined so that the detailed balance
condition holds for the overall updating procedure and the sum rule
\begin{equation}
  \sum_{\psi_b} p(\psi_b|\phi_b)=1
\label{renorm}
\end{equation}
is satisfied. Generally, this assignment is an over-determined
problem, but in the present problem, we can assign the probabilities
uniquely, as we now discuss.

To define these probabilities, we first define a new local weight
\beq
  \tilde w_{\psi_b}(\phi_b) \equiv c_{\psi_b} w(\phi_b) p(\psi_b | \phi_b),
  \label{prob}
\eeq
where $c_{\psi_b}$ is a configuration independent renormalization
constant which is to be determined so that \Eq{renorm} holds.  In the
second step of the algorithm, the stochastic updating of states, a
Monte Carlo step is performed using (\ref{prob}) instead of
(\ref{localw}).  As long as the relationship \Eq{prob} holds, this
step together with the first step constitutes a single Monte Carlo
move which satisfies the detailed balance condition \cite{kawashima94a}.

Following the viewpoint of Kawashima and Gubernatis
\cite{kawashima94a}, the algorithm is characterized by the modified
weights $\tilde w_{\psi_b}(\phi_b)$ since once we specify them and
since the $w(\phi_b)$ are given, the probabilistic assignment of the
labels, i.e., $p(\psi_b | \phi_b)$, is determined by \Eq{renorm} and
\Eq{prob}. To show this, we define
\begin{equation}
 \tilde w_{\psi_b}(\phi_b)=\tilde u_{\psi_b}(\phi_b^\uparrow)
                           \tilde u_{\psi_b}(\phi_b^\downarrow)
                            v(\phi_b^\uparrow,\phi_b^\downarrow)
\end{equation}
where $v(\phidn_b,\phiup_b)$ is the same as in \Eq{localw} and
\begin{equation}
   \tilde u_x(y) = \left\{
          \begin{array}{ll}
           0, & {\ \rm if\ } x = y {\rm\ or\ } x = {\bar y}\\
           1, & {\ \rm otherwise}
          \end{array}
                    \right.
\label{eq:ModWgt}
\end{equation}
The probability for the labeling procedure can then be written as
\beq
  p(\psi_b|\phi_b) = p_{\rm P}(\psidn_b|\phidn_b) p_{\rm P}(\psiup_b|\phiup_b).
\eeq
For a given spin, the labeling probability $p_P$ of one plaquette is
easily found to be
\beqa
    p_{\rm P}(2|1) &=& 1-p_{\rm P}(3|1)
     = p_{\rm P}(2|\bar 1) = 1-p_{\rm P}(3|\bar 1) = P_1, \nonumber \\
    p_{\rm P}(3|2) &=& 1-p_{\rm P}(1|2)
     = p_{\rm P}(3|\bar 2) = 1-p_{\rm P}(1|\bar 2) = P_2, \nonumber \\
    p_{\rm P}(1|3) &=& 1-p_{\rm P}(2|3)
     = p_{\rm P}(1|\bar 3) = 1-p_{\rm P}(2|\bar 3) = P_3,
\eeqa
where
\begin{equation}
    P_1  =  \frac{1}{2}(1-\e^{-\tau t}), \quad
    P_2  =  \frac{1+\e^{\tau t}}{1+\e^{2\tau t}},\quad
    P_3  =  \frac{1-\e^{-\tau t}}{1-\e^{-2\tau t}}
\end{equation}

In order to make the simulation effective, it is desirable to have an
updating rule which guarantees that any resulting configuration is an
allowed state.  This is easy to do in the present case, since the
state which can be reached by flipping a loop configuration is an
allowed one.  Actually, the modified weights (\ref{eq:ModWgt}) are
chosen so that this correspondence holds.  In order to explain what
loop configurations are, we define how we construct a loop: For each
shaded plaquette, we connect each corner site to another corner by a
segment as depicted in \Fig{fg:L}. The connected sites form loops.  We
note that any configuration reached by flipping a loop is allowed
since a flip of any segment of a loop results in an allowed local
configuration.  {\it Flipping a loop} means replacing all the
electrons on the loop by holes and vice versa.

%Note: Deleted some lines that  already appear a few lines earlier.
%
If the $v$ term in \Eq{prob} were not present, we would simply flip
each loop independently with probability $1/2$.  In our case, with the
$v$ term present, each loop is flipped with a probability
\beq
      R \equiv 1/(1+\e^{-\tau\Delta})
\label{eq:accept_ratio}
\eeq
where
\beq
 \Delta \equiv  \frac{U}{2}(2{\cal M-L})-\frac{\mu}{2}(2{\cal N-L}),
\label{eq:loopw}
\eeq
where ${\cal N}$ is the total number of particles on the loop, ${\cal
M}$ is the total number of sites on which there are zero or two
electrons, and ${\cal L}$ is the length of the loop.  We remark that the
flip of one loop can change the value of $\Delta$ for the other loops.
We, therefore, flip loops sequentially, one after the other.  For $N=2$
and $L=4$, an example of a loop-flip is shown in Fig.~\ref{fg:5}.

We remark that any change of the worldlines by algorithm \algP\
can be realized by algorithm \algL\ with a finite probability.
Additionally, any winding number can occur, and thus the algorithm
simulates the grand canonical ensemble of the model.  We can prove
ergodicity and even a stronger statement that any state which does not
violate local particle-number conservation can be reached with a
finite probability from any other such state in one sweep of Monte
Carlo updating. These statements are a consequence of a loop flip
corresponding to changing the occupancy on all sites connected by
the loop.  Conversely, the difference between a pair of allowed
worldlines configurations corresponds exactly to a loop configuration.
Thus, any two allowed worldline configurations can be transformed in
to each other by a single flip of a loop configuration.

%%%%%%%%%%%%%%%%%%%%%%%%%%%%%%%%%%%%%%%%%%%%%%%%%%%%%%%%%%%%%%%%%%
%%%%%%%%   L o o p   E X C H A N G E   A L G O R I T H M   %%%%%%%
%%%%%%%%%%%%%%%%%%%%%%%%%%%%%%%%%%%%%%%%%%%%%%%%%%%%%%%%%%%%%%%%%%
\subsection{The loop-exchange algorithm (Algorithm \algLex)}
\label{subsec:Lex}

The loop-flip algorithm described above was an extension of the
algorithm that Evertz et al.\ applied to the 6-vertex model; however,
for the Hubbard model it has shortcomings which do not exist for the
6-vertex model or a quantum spin system.  In particular, when $U$ is
large compared to $T$, long loops have little chance of being flipped
\cite{somsky92}.  We can understand this behavior by observing that
long loops decrease the acceptance probability (\ref{eq:loopw}) when
$U>0$.  Since we are often interested in the model with large positive
$U$ and small $T$, this can be problematic as it can lead to long
auto-correlation times.  In addition, we are also interested in
non-half-filled cases.  For them, we also face a similar difficulty of
long correlation times because flipping a loop may change the total
number of particles, and the acceptance ratio (\ref{eq:accept_ratio})
may therefore be strongly be suppressed by the $\mu$-term
(\ref{eq:loopw}).

In order to resolve these difficulties, we developed a new loop
algorithm, the loop-exchange algorithm (algorithm \algLex).  In this
algorithm, two loops which have the same shape, but different spins,
are flipped simultaneously.  In other words, these flips do not change
the coefficients of $U$-term nor $\mu$-term in (\ref{eq:loopw}).
Inherently, this algorithm is non-ergodic so it must be used with some
ergodic algorithm to construct a correct Markov process.  In the present
paper, we use algorithm \algL\ for this purpose.

We will discuss algorithm \algLex\ as a modification of algorithm
\algL. In algorithm \algLex, all 36 configurations of a block, {\it
except} two, map uniquely onto a smaller set of labels.  The rule for
label assignment is straightforward: First, we draw worldlines by
following the ordinary rule described in \ssc{subsec:Worldlines}.
Then, we overlap these two plaquettes and erase all doubly drawn
lines, i.e., worldline segments drawn at the same place on both the
plaquettes.  If the resulting picture has lines that dead-end at
opposite corners of the same plaquette, these corners are connected by
a straight line. After doing this for all blocks, we identify the
resulting sets of overlapping worldline segments with labels and
assign these labels with a probability of one.  This labeling is
illustrated in Fig.~\ref{fg:Lex}.

The two exceptional cases are those for which the rule of the label
assignment is not unique.  These cases are configurations $(2,\bar 2)$
and $(\bar 2, 2)$.  Each can have one of two labels, which we denote
by 1 and 2.  Their modified weights are defined as
\begin{equation}
     \tilde w_1((2, \bar 2)) = \tilde w_1((\bar 2, 2)) =
     \tilde w_2((2, \bar 2)) = \tilde w_2((\bar 2, 2)) = 1
\end{equation}
with all other modified weights, i.e., those $\tilde w_i$ with the two
exceptional labels but with $i$ not equal to 1 or 2, vanishing.

With the specification of all other modified weights as unity, the
definition of the loop exchange algorithm is now complete.  We also
note that once the labels are chosen, the modified weights for all
allowed loop configurations are equal and the transition probability
of a Monte Carlo attempt is one-half for any loop, no matter how large
$U$ and $\mu$ may be.  In terms of labeling probabilities, the
algorithm
\algLex\ is characterized by
\beqa
    p(1|(2,\bar 2)) &=& p(1|(\bar 2,2)) = 1 - \th^2 \tau t\\
    p(2|(2,\bar 2)) &=& p(2|(\bar 2,2)) = \th^2 \tau t.
\eeqa
For the other configurations, the labeling probability is 1 or 0 as
described above.

For algorithm \algLex, only an allowed state can be reached from a
flipped loop.  Both the up and down planes are treated simultaneously,
whereas in algorithm \algL, they are treated independently as far as
the loop construction is concerned.  In algorithm \algLex, therefore,
only one set of loops exists for the two planes.  The construction
rule for the loops is simple: All we do is just connect the segments
defined above, which have a one-to-one correspondence to the labels.
This rule gives connected loops; no open curves appear.  A useful way
to view this is as follows: The plaquettes in the right column of
Fig.~\ref{fg:Lex} are the result of an exclusive-or (XOR) operation on
the overlaid plaquettes along the rows to the left.  This XOR
operation replaces the overlaid worldlines by a lattice with 0's of
1's at the lattice sites. The \algLex\ loops connect the sites
with the 1's, of which there are 0, 2, or 4 per shaded plaquette.
Thus, the loops are divergence-free.  An example of a loop-exchange is
shown in Fig.~\ref{fg:7} when $N=4$ and $L=3$.

%%%%%%%%%%%%%%%%%%%%%%%%%%%%%%%%%%%%%%%%%%%%%%%%%%%%%%%%%%%%%%%%%%
%%%%%%%%%%%%%%%%%%%%%%%   R E S U L T S  %%%%%%%%%%%%%%%%%%%%%%%%%
%%%%%%%%%%%%%%%%%%%%%%%%%%%%%%%%%%%%%%%%%%%%%%%%%%%%%%%%%%%%%%%%%%
\section{Results}
\label{sec:results}

In this section, we present our numerical results for the integrated
auto-correlation times (\ref{eq:TauBin}) associated with the average
energy, electron occupancy (\ref{eq:N}), and the $q=\pi$, static
charge-density $T\chi_{+}$ and spin-density $T\chi_{-}$ correlations
(\ref{eq:Chi}).  We will denote these times by
$\tau_{int}^E$, $\tau_{int}^N$, $\tau_{int}^C$, and $\tau_{int}^S$. We
will also present results for $\tauexp$.

In what follows, we will be concerned with various combinations of the
three algorithms described in the previous sections.  One case is the
plaquette algorithm \algP\ by itself.  Another is the combination of
\algP\ and \algL, which we will refer to as $PL$.  Other symbols, such
as $L$, $LL_{\rm ex}$, and $PLL_{\rm ex}$, should be understood in a
similar fashion.  The application of any combination which includes
$L$ results naturally in a grand canonical and ergodic simulation,
while the application $P$ and $L_{\rm ex}$ alone or together results
in a canonical and non-ergodic simulation.  Although we can force
algorithm $L$ to simulate the canonical ensemble, we did not do this
unless otherwise stated.

As we mentioned above, in some cases, especially when algorithm $P$
was used alone, we were unable to estimate $\tauint^A$ because the
$\tauint^A(l)$ versus $l$ curve did not reach a plateau during the
course of our rather lengthy Monte Carlo runs.  In such cases, we took
the largest available value (with still a reasonably large number bins
remaining) as a practical lower bound on $\tauint^A$.  This choice of
a lower bound is justified since all the $\tauint^A(l)$ versus $l$
curves we computed, whether they reached the plateau or not, were
non-decreasing functions.  In \Fig{fg:tauvsl}, we show
$\tauint^A(l)$ versus $l$ curves for a typical example where using
algorithm $P$, we were forced to take a lower bound, and where using
algorithm $PLL_{\rm ex},$ we reached a plateau.  This figure also
illustrates that in some cases we can reduce the autocorrelation time
dramatically by combining $P$ with $L$ and $L_{\rm ex}$.

For most simulations, the length of the Monte Carlo calculation was
0.25 million Monte Carlo steps (MCS).  In some exceptional cases, where the
autocorrelation times are very long, we performed longer runs
\cite{cases}. In algorithm \algP, a Monte Carlo step consists of a
sweep across all unshaded plaquettes on which an attempt to flip its
corners is made.  In algorithms \algL\ and \algLex, a Monte Carlo step
means decomposing the whole system into loops and attempting to flip
every loop.  When several algorithms are combined, an MCS consists of
one sweep through each algorithm.  For example, in $PL$, one Monte
Carlo step is one sweep through $P$ and one sweep through \algL.

In \Fig{fg:tau}, various autocorrelation times for $P$ and $PLL_{\rm
ex}$ are plotted for fixed value of $y \equiv U\beta/L = 0.5$,
where $L$ is the lattice extension in imaginary time direction.
 From \Fig{fg:tau}(a), (b), and (c), we see that for our new algorithm
\PLLex, the autocorrelation time is very small, of order one, in
almost all cases. After some initial increase, it is flat as a
function of $1/T$ and also as a function of $U$.  Thus, algorithm
\PLLex\ performs very well.

The behavior of the algorithm $P$, the conventional algorithm, is not
as clear.  At small $\beta$, where we have performed very long
simulations, we see that $\tauint$ increases rapidly as a function of
$\beta$.  We also notice also that as a function of $U$, $\tauint$
increases rapidly.  At larger $\beta$, most of our (fairly short) runs
with algorithm $P$ did not converge, and we show only lower bounds for
$\tauint$.It is possible that the initial increase of $\tauint$ with
$\beta$ actually continues towards larger $\beta$.  Therefore, some of
the lower bounds on $\tauint$ could possibly be an order of magnitude
below the actual value.

In short, comparing $P$ and \PLLex, we see that at large $U$ and small
$\beta$, the difference in performance between these algorithms is
enormous, being more than 3 orders of magnitude at $U=8$.  As argued
above, there could be an even bigger difference in performance when
$\beta$ becomes large.

Figure~\ref{fg:tau} also shows that our $PLL_{\rm ex}$ algorithm is
very effective in reducing the integrated correlation time for the SDW
correlation function.  The reason for this is that we can exchange
worldlines for an up-spin and a down-spin electron in either algorithm
$L$ and $L_{\rm ex}$ without visiting intermediate states with small
weights; however, algorithm $L_{\rm ex}$ is most effective in this
regard as it was designed to do precisely this. At large $U$ and at
low temperature, the typical spatial worldline configuration is the
one where up-spin and down-spin worldlines appear alternatively with
small overlap.  The overlap is strongly discouraged by a large value
of $U$.  In algorithm $P$, in order to exchange the positions of two
worldlines, we need intermediate states in which the two worldlines
overlap at least partially.  We avoid these intermediate states with
the algorithms $L$ and $L_{\rm ex}$.  We have observed strong SDW
ordering of the system by looking at the average values of $T\chi_{-}$
at $q=\pi$. To show an example, $T\chi_{-}(\pi) = 0.884(1)$ at $U=8$
and $\beta=0.5$ for $PLL_{\rm ex}$ whereas it becomes $T\chi_{-}(\pi)
= 2.000(8)$ at $\beta=8$.

Figure~\ref{fg:tau}(c) for the CDW function sharply contrasts that of
Fig.~\ref{fg:tau}(b) for the SDW function in that a significant
increase in the correlation time does not occur as the temperature is
lowered.  (We again emphasize that the $\tau_{int}$ displayed for
algorithm P are in general lower bound estimates and are likely to be
too small.) In fact, the correlation time is almost constant for
$\beta > 6$ in all the six curves.  The difference between $P$ and
$PLL_{\rm ex}$ is far less noticeable: the correlation time for $P$ is
only a few times larger than its counterpart for $PLL_{\rm ex}$.
Consistent with these observations is the fact that CDW order is much
weaker than the SDW order.  For example, $T\chi_{+}(\pi) = 0.1281(5)$
at $U=8$ and $\beta=0.5$ for $PLL_{\rm ex}$ whereas it becomes
$T\chi_{+}(\pi) = 0.00380(2)$ at $\beta=8$.

The longest integrated correlation times for $PLL_{\rm ex}$ are seen
in \Fig{fg:tau}(d) for the average electron number.  Since there are
no fluctuations in the particle number in algorithm $P$, there is no
corresponding curves for $P$ for this quantity.  However, even for
$PLL_{\rm ex}$, the fluctuation in the particle number is very small.
The variance in the particle number is sometimes so small that we
could not obtain a reliable estimate for its correlation time. This
undetermination is why some data points are missing in the figure,
especially in the region of large correlation times at lower
temperatures.

The comparison between $\tauint$ and $\tauexp$ in \Fig{fg:tau}(d) shows
that in most cases the integrated correlation time $\tauint$ for the
particle number is nearly as large as the longest mode correlation
time $\tauexp$. This fact suggests that the longest mode in $PLL_{\rm
ex}$ is the mode in which creation and annihilation of particles are
involved.  Empirically, the small variance in the particle number and
the long correlation time associated with charge creation and
annihilation appear closely related.  For example, in the case where
$N=32$, $U=8$, $\mu=0$, $1/T=4$ and $N=32$, the square-root of the
particle number variance (i.e., one standard deviation) measured by
the algorithm $PLL_{\rm ex}$ is 0.02.  For lower temperatures, with
the rest of parameters the same, we were unable to estimate this
quantity reliably.  However, the influence of this long mode for
relevant physical quantities may be  negligible because the change in
this quantity may be so small that we can effectively consider it as a
constant.  This is the case with the long mode associated to the
average particle number, since the average values of other physical
quantities, such as energy and static SDW susceptibility, are not as
sensitive to particle number fluctuations so this exponentially small
fluctuation cannot appreciably affect the average value of those
quantities.  In this sense, the longest mode correlation time is not
necessarily the relevant measure for the computational time, except if
we are interested in the longest mode itself.

On the other hand, although the creation and annihilation of particles
are rather rare and cause long $\tauexp$'s in some cases, we found
that it is still important in other cases to allow them to happen in
order to reduce the integrated correlation times for quantities of
physical interest.  To see this, we have performed some simulations in
which all the Monte Carlo attempts for loops whose winding number is
non-vanishing (and therefore may change the particle number) are
rejected.  In \tab{tb:GCvsC}, we show the correlation times for these
special simulations in comparison with the grand canonical
simulations.  The result clearly shows the importance of creation and
annihilation processes.

As already mentioned, the present algorithm allows negative sign
configurations to occur.  We found, however, the negative sign ratio
defined by \cite{loh90}
\beq
    \Gamma \equiv (Z_+ - Z_-) / (Z_+ + Z_-)
\eeq
where $Z_+$ ($Z_-$) is the Boltzmann weight for the subset of
configurations that are positive (negative), is close to unity for
most cases, especially for larger system sizes.  This observation
means that the ``sign problem'' is not a problem for the new algorithms.
In fact, the only cases for which we found any configurations of
negative sign are the cases where $N=32$, $\mu=0.0$, $U=2$, $y=0.5$,
and $\beta=4$, $6$, $8$, and $10$. The values of $\Gamma$ for these
cases are $0.9978(6)$, $0.979(4)$, $0.964(5)$ and $0.88(1)$.  At
larger values of $U$ or higher temperatures, we did not observe any
negative sign configurations.  For smaller lattices, such as $N=16$,
we observed a larger number of negative sign configurations, although
they are again just a small fraction of the total.  We expect that the
negative sign ratio approaches to unity as the system size becomes
large.

The reduction in auto-correlation times brought about by algorithm
$L_{\rm ex}$ is most noticeable for the integrated correlation time
for the SDW correlation function.  In \tab{tb:algorithms}, we list the
correlation times for various combinations of the three algorithms, in
the case where $U=4$, $\beta=1.5$, $N=32$ and $\mu=0.0$, to examine
the efficiency of several combinations.  We can see that the
integrated correlation time $\tauint^{S}$ for the SDW correlation
function for \PLLex\ is about 1/8 of that for $PL$.  Since we have not
carried out this type of investigation for a variety of physical
parameters systematically, cases may exist where the efficiency of
$L_{\rm ex}$ is much more noticeable than presented here.  The reason
why the efficiency of $L_{\rm ex}$ is most noticeable in the
correlation time for SDW is conjectured as follows: As already argued,
the integrated correlation time for SDW fluctuations is related to the
exchange of two worldlines of opposite spins.  Although we can achieve
this exchange through either algorithm $P$ or $L$, we inevitably
encounter unpleasant intermediate states in the case of $P$.
Algorithm $L$ may suffer from similar difficulty because in forming
loops in either the up-spin plane or the down-spin plane it does not
use the information about the configuration in the other plane.
Therefore, in some cases, the attempted movement of the worldline
results in larger overlaps between down-spin worldlines and up-spin
ones, and the move is improbable. On the other hand, the algorithm
$L_{\rm ex}$ is reasonably free from this kind of difficulty.

The acceptance ratios of the various algorithms also shows a wide
range of behavior.  In \Table{tb:AcceptanceRatio}, we list the
acceptance ratios for some typical cases. The acceptance ratio $r$ for
one MCS is defined as
\beq
  r \equiv \frac{\mbox{[The number of flipped sites]}}
                {n \times \mbox{[The number of sites in the system]}},
\eeq
where $n=2$ for $P$ and $n=1$ for $L$ and $L_{\rm ex}$.  The integer
$n$ is the number of times a site is flipped in a MCS.  For the
algorithm $P$, this definition is equivalent to the ordinary
acceptance ratio defined as the number of accepted attempts divided by
the total number of attempts.  The numbers in
\Table{tb:AcceptanceRatio} are averaged over all the Monte Carlo
steps.  We see that the acceptance ratio for $P$ and $L$ are strongly
dependent on $\beta$ and $U$ and become very small as the value of $U$
increases.  In contrast, the acceptance ratio for $L_{\rm ex}$ is much
less parameter dependent and only moderately increases to $1/2$ as $U$
increases.

Some results on the quarter-filled case are presented in
\tab{tb:quarterfilling}.  In this case, in order to compare the
results to those of the algorithm $P$ on a roughly equal basis, we
adjusted the chemical potential $\mu$ for the algorithm $PLL_{\rm ex}$
so that the resulting average particle number becomes close to 0.5.
The difference between the algorithms is less striking than that in
the half-filled case although we can still see considerable reduction
in the correlation time.  We remark that for $PLL_{\rm ex}$ the
autocorrelations time for 1/4-filling are comparable to those at
1/2-filling. Since the overhead of $PLL_{\rm ex}$, relative to $P$, is
approximately a factor of 10 larger, the quarter-filled case is near a
marginal filling where the overhead and the reduction of correlation
time balances.

%%%%%%%%%%%%%%%%%%%%%%%%%%%%%%%%%%%%%%%%%%%%%%%%%%%%%%%%%%%%%%%%%%
%%%%%%%%%%%%%%%%%%%%   C O N C L U S I O N   %%%%%%%%%%%%%%%%%%%%%
%%%%%%%%%%%%%%%%%%%%%%%%%%%%%%%%%%%%%%%%%%%%%%%%%%%%%%%%%%%%%%%%%%
\section{Conclusion}
\label{sec:conclusion}

We presented two new algorithms for worldline Monte Carlo simulations
of electron systems and demonstrated their efficiency in the case of
the one-dimensional, repulsive Hubbard model.  Their extension to
higher dimensions is straightforward. The first algorithm, which we
call the loop-flip algorithm, is an extension to fermion systems of a
loop-flip algorithm for 6-vertex model recently proposed by Evertz et
al. \cite{evertz93}. Our new algorithm enables us to simulate the
grand canonical ensemble while the traditional plaquette flip
algorithm simulates in the canonical ensemble.  We also found that
switching to grand canonical simulation often reduces the
autocorrelation time of the simulation dramatically.  The second
algorithm is the loop-exchange algorithm in which up-spin and
down-spin worldline segments exchange their locations.  Since all
worldline movement in the loop exchange algorithm is unaccompanied by
changes in the chemical potential term or the $U$-term in the
Hamiltonian, we expected that this algorithm would be effective when
the long correlation time is due to difficulties in exchanging
worldlines.  Consistent with this expectation, we found that the loop
exchange algorithm is especially effective in reducing the
autocorrelation time of the SDW correlation function when the value of
$U$ is large.  These algorithms generalize the cluster algorithms
\cite{kandel91} recently developed mainly to reduce long
auto-correlation times accompanying simulations of critical phenomena
in classical spin systems.  In contrast to the usual use of cluster
algorithms, we were concerned with reducing the inherently long
auto-correlation times that occur in the worldline method even when
the physical system is far removed from any known finite-temperature
phase transition.

At the lowest simulated temperatures, we observed reductions of the
autocorrelation time, measured in the units of Monte Carlo steps, as
large as three orders of magnitudes.  One MCS in the algorithm
$PLL_{\rm ex}$, however, takes from 8 to 10 times longer in CPU time
than one step in algorithm $P$ so the above reduction of
autocorrelation time gives a reduction in CPU time by a factor as
large as two orders of magnitude.

The estimates of improved efficiency are cautious ones. In most cases
we were able only to estimate a lower bound for the autocorrelation
times associated with the plaquette algorithm. It is not unreasonable
in some casses to expect the actual autocorrelation times to be an
order of magnitude larger. Additionally, it should be possible to
improve the efficiency our our implementation of the loop algorithms.
We used a highly optimized code for the plaquette algorithm but used
unoptimized codes for the loop algorithms.  Further, we used ordinary
workstations for all the calculations.  Parallelizing these
computations is possible and will most likely be needed when the
present algorithms are applied more challenging problems, such as
those in higher dimensions.

It is natural to ask about the extensions of the present methods to
higher dimensions and other models. When one changes the problem, it
is perhaps best to ask whether one should also change the methods.
The algorithms presented were designed for the one-dimensional
repulsive Hubbard model but can be viewed from the perspective of a
general approach to developing cluster algorithms
\cite{kawashima94a}.  Tailoring the algorithms, when extended to other
problems, might be possible.  Still, the direct use of our new methods
to many other problems is possible.  How efficient will be this use is
what needs investigation.  For many cases, their use should be more
efficient than the standard algorithm.

The most obvious extension would be to the two-dimensional repulsive
Hubbard model.  The standard worldline algorithm suffers such severe
sign problems in two-dimensions that is often almost pointless to use
it.  We have no reason to believe our algorithms reduce the sign
problem.  Our best expectation would be the achievement of a
significant enough reduction in the variance of the measured physical
quantities so simulations could be performed over a limited range of
parameters.  Wiese \cite{wiese93a} has reported the use an extension
of the massless version of the Evertz et al.\ loop-flip algorithm on
the two-dimensional, free fermion problem, and the few high
temperature results he reported show seemingly accurate values for the
average electron occupancy even in the presence of a severe sign
problem.  When $U=\mu=0$ our loop-flip algorithm for fermions reduces
to his if we only attempt to flip one loop at each Monte Carlo step.
The potential effectiveness of these methods for fermions problems in
two or more dimensions, while appearing not especially promising,
needs more study.

Of our two algorithms, the loop-exchange method is the one most
specific to the repulsive Hubbard and related models.  If $U<0$, an
algorithm, which breaks-up worldines wanting to be on top of one
another, would be more appropriate and should be possible to
construct. For the extended Hubbard model, which has a Coulomb term
between electrons on neighboring sites, an additional specific
construct might also be necessary.  The loop-exchange algorithm,
because is addresses the hopping part of the problem, will change
little from problem to problem.

Loop algorithm $L$, in fact, can be almost directly applied to the
spin-1/2 quantum spin chains.  This applicability follows from the
similarity the path-integral representation of the spin-1/2 quantum
spin system has with spinless fermion model \cite{somsky92}.  Use of
these algorithms in two-dimensions also appears quite direct and
desirable.  Wiese and Ying \cite{wiese93b}, using what turns out to be
the massless version of of the Evertz et al.\ method, studied the
spin-1/2 Heisenberg anti-ferromagnet with success, although they do
not report efficiency figures of merit.  Recently, we developed
cluster algorithms for spin of arbitrary magnitude \cite{kawashima94b}
and are in the process of testing these methods. We will report our
results elsewhere.

In closing, we comment that the total temporal and spatial winding
numbers are both physically important.  The total temporal winding
number is the total number of particles and the average of the square
of its fluctuations is related to the charge compressibility of the
system.  The average of the square of the fluctuations of the total
spatial winding number is directly related to the dc conductivity
\cite{scalapino}.  In the conventional worldline Monte Carlo, i.e.,
algorithm $P$, the total winding number of either type is conserved.
To estimate the charge compressibility and dc conductivity, we have to
do some additional manipualtions.  In the new algorithms, these
manipulations seem unnecessary because the total winding numbers of
both kinds can change simulaneously, individually, or not at all.  We
can see the changes by noting that in algorithm $L$ flipping a
single loop of spatial winding number $w$ changes the total spatial
winding number by $w$, and that a loop with any spatial winding number
can form with a finite probability as long as it is allowed in the
size of the system under consideration.  A similar remark can be made
with respect to the temporal winding number. Therefore, we can
directly estimate the above-mentioned quantities.  We have yet to
investigate the efficiency of this possibility because in this paper
we were mainly concerned with overall algorithmic issues and
development.  An additional important point is that similar
possibilities will also exist for the relevant extension of the loop
algorithms to bosons and quantum spin systems.

\acknowledgments

The work of J.E.G.\ was supported by the Department of Energy's High
Performance Computing and Communication program at the Los Alamos
National Laboratory. H.G.E.\ would like to express his gratitude to
the Department of Energy for providing computer time on the Cray Y-MP
computer at Florida State University. We thank A.\ Sandvik for several
helpful coments regarding our original manuscript.

%%%%%%%%%%%%%%%%%%%%%%%%%%%%%%%%%%%%%%%%%%%%%%%%%%%
%%%%%%%%        R E F E R E N C E S        %%%%%%%%
%%%%%%%%%%%%%%%%%%%%%%%%%%%%%%%%%%%%%%%%%%%%%%%%%%%

%%%%%%%%%%%%%%%%%%%%%%%%%%%%%%%%%%%%%%%%%%%%%%%%%%%
%%%%%%%%            T A B L E S            %%%%%%%%
%%%%%%%%%%%%%%%%%%%%%%%%%%%%%%%%%%%%%%%%%%%%%%%%%%%

\begin{table}
\caption{Auto-correlation times for $U=8$, $L=8$, $1/T=1$, $N=16$,
and $\mu=0$.}
\label{tb:GCvsC}
\ \\
\begin{tabular}{lcccccc}
Algorithm& $\tauint^{E}$ & $\tauint^{N}$ & $\tauint^{C}$ & $\tauint^{S}$ &
$\tauexp$\\
\hline
\PL\ (canonical)      & $>$46  &$\infty$ & 13(2)   & 60(5)   & ---$^{\rm a}$\\
\PL\ (grand canonical)& 1.7(2) & 0.89(3) & 0.81(1) & 10.8(9) & 19(7) \\
\PLLex\ (canonical)   & $>$28  &$\infty$ & 10.7(1) & 1.80(5) & ---$^{\rm a}$\\
\PLLex\ (grand canonical)& 0.99(4)& 0.92(2) & 0.718(7)& 0.509(8)& 1.75(5) \\
\end{tabular}
${}^{\rm a}$ Unable to measure.
\end{table}

\begin{table}
\caption{Auto-correlation times for $U=4$, $L=12$, $1/T=1.5$, $N=32$, and
$\mu=0$.}
\label{tb:algorithms}
\ \\
\begin{tabular}{lccccc}
Algorithm & $\tauint^{E}$ & $\tauint^{N}$ & $\tauint^{C}$ & $\tauint^{S}$ &
 $\tauexp$ \\
\hline
$P$    & 102(7) &$\infty$ & 17(1)   & 45(2)   & ---$^{\rm a}$\\
$L$    & 4.2(3) & 1.22(6) & 0.73(3) & 5.6(5)  & 10(1)       \\
\PL    & 2.0(2) & 1.20(3) & 0.63(2) & 3.8(1)  & 5.1(5)      \\
\PLex  & $>$33  &$\infty$ & $>$5.2  & $>$1.7  & ---$^{\rm a}$\\
\LLex  & 3.1(2) & 1.28(4) & 0.711(4)& 0.60(3) & 5.9(8)      \\
\PLLex & 1.26(2)& 1.28(8) & 0.63(3) & 0.525(4)& 3.4(6)      \\
\end{tabular}
${}^{\rm a}$ Unable to measure.
\end{table}

\begin{table}
\caption{Acceptance ratios for $N=32$, $U/\beta L=1$, and $\mu=0$.}
\label{tb:AcceptanceRatio}
\ \\
\hfill
\begin{tabular}{lcccc}
 $\beta$& Algorithm    & $U=2$ & $U=4$ & $U=8$ \\
\hline
 2      & $P$          & 0.146 & 0.071 & 0.020 \\
        & $L$          & 0.208 & 0.074 & 0.012 \\
        & $L_{\rm ex}$ & 0.349 & 0.413 & 0.478 \\
\hline
 6      & $P$          & 0.157 & 0.081 & 0.028 \\
        & $L$          & 0.105 & 0.050 & 0.014 \\
        & $L_{\rm ex}$ & 0.355 & 0.415 & 0.471 \\
\end{tabular}
\hfill
\end{table}

\begin{table}
\caption{Auto-correlation times for $N=16$, $L=8$, and $U=8$,
at quarter filling or nearly quarter filling. }
\label{tb:quarterfilling}
\ \\
\begin{tabular}{lccccccc}
Algorithm&$1/T$&$\la n
\ra$&$\tauint^{E}$&$\tauint^{N}$&$\tauint^{C}$&$\tauint^{S}$&$\tauexp$\\
\hline
    $P$ & 1.0 & 0.5 & 12.4(9) & $\infty$ & 6.8(4)  & 9.9(5) & ---$^{\rm a}$\\
    $P$ & 2.0 & 0.5 & 22(5)   & $\infty$ & 8.1(3)  & 10.6(8)& ---$^{\rm a}$\\
    $P$ & 4.0 & 0.5 & 28(2)   & $\infty$ & 10.0(7) & 11(1)  & ---$^{\rm a}$\\
%   $P$ & 6.0 & 0.5 & 23(3)   & $\infty$ & 10.7(7) & 7.9(4) & ---$^{\rm a}$\\
 \PLLex & 1.0 & 0.4975(2) & 1.12(5) & 1.24(6) & 0.55(1)  & 0.514(5) & 1.21(3)\\
 \PLLex & 2.0 & 0.4940(6) & 1.51(5) & 3.3(2)  & 0.786(9) & 0.60(2)  & 4.0(1) \\
 \PLLex & 4.0 & 0.4824(3) & 2.0(2)  & 34(4)   & 1.49(3)  & 1.1(1)   & 35(2)  \\
%\PLLex & 6.0 & 0.518(1)  & 3.2(2)  & 180(10) & 1.8(1)   & 2.5(1)   & $>$130 \\
\end{tabular}
${}^{\rm a}$ Unable to measure.
\end{table}

%%%%%%%%%%%%%%%%%%%%%%%%%%%%%%%%%%%%%%%%%%%%%%%%%%%
%%%%%%%%          C A P T I O N S          %%%%%%%%
%%%%%%%%%%%%%%%%%%%%%%%%%%%%%%%%%%%%%%%%%%%%%%%%%%%

\begin{figure}
\caption{
An example of the space-time checkerboard for a single component of
spin. Here, 4 electrons of the same spin are on 8 lattice sites, which
are along the horizontal direction.  In the vertical direction are
$2L=8$ imaginary-time steps.  The symbols $S_i$ along the left edge
designate the many-body state at that imaginary-time step.  The ket
$|S_2\rangle = |1,1,0,0,1,1,0,0\rangle$, for example.
}
\label{fg:1}
\end{figure}

\begin{figure}
\caption{
The six allowed shaded plaquettes. The values of $\phi$ are our symbols
for the depicted plaquette.
}
\label{fg:2}
\end{figure}

\begin{figure}
\caption{
The allowed movements of worldline segments back and forth across a unshaded
plaquette.
}
\label{fg:3}
\end{figure}

\begin{figure}
\caption{
The labeling probability for the algorithm \algL.  In the top row, the
correspondence between labels and loop decompositions is shown.  In
the leftmost column, the correspondence between symbols ($1,\bar1,
2,\bar 2, 3$ and $\bar 3$ ) and states are shown. Solid circles stand
for particles while open circles for holes.
}
\label{fg:L}
\end{figure}

\begin{figure}
\caption{
An example of a loop flip: (a) a configuration with a single
fermion, (b) a possible decomposition resulting in three loops, and
(c) a possible worldline configuration resulting from the loop flips.
Depicted in (c) is a two fermion configuration with a temporal winding number
of two and a spatial winding number of one.  Between (a) and (c), a sign
change occurred.
}
\label{fg:5}
\end{figure}

\begin{figure}
\caption{
Correspondence between local states and loop segments in the algorithm
\algLex.  The states are depicted in terms of worldline segments.
Solid lines are for down-spin particles and dashed lines are for
up-spin particles.
}
\label{fg:Lex}
\end{figure}

\begin{figure}
\caption{
An example of a loop-exchange.
}
\label{fg:7}
\end{figure}

\begin{figure}
\caption{
The bin-length length dependence of $\tau_{int}^A(l)$ for the SDW fluctuation
for both the $P$ and $PLL_{\rm ex}$ algorithms. (a) A case where
$\tau_{int}^A(l)$ does not converge ($N=32$, $L=160$, $U=8$,$\beta=8$ and
$\mu=0.0$). (b) A case where  $\tau_{int}^A(l)$ converges.
}
\label{fg:tauvsl}
\end{figure}

\begin{figure}
\caption{
The integrated auto-correlation times for (a) energy, (b) SDW
fluctuation, (c) CDW fluctuation and (d) the number of particles, in
the case of $N=32$, $\mu = 0.0$ and $U\beta/L =0.5$.  The double
symbols and dashed curves show (most likely poor) lower bounds for
the correlation times, not the correlation times themselves.  The
curves are computed for several values of $U$.
}
\label{fg:tau}
\end{figure}


\begin{references}

\bibitem{suzuki87} For examples of applications to quantum spin
systems, see {\it Quantum Monte Carlo in Equilibrium and
Non-equilibrium Systems}, edited by M.~Suzuki
(Springer-Verlag-Heidelberg, 1987).

\bibitem{negle88} J.W.~Negle and H.~Orland, {\it Quantum Many-Particle
Systems}, (Addison-Wesley, New York, 1988), Chap.~8.

\bibitem{hirsch81} J.E.~Hirsch, R.L.~Sugar, D.J.~Scalapino, and
R.~Blankenbecler, Phys.\ Rev.\ Lett.\ {\bf 47}, 1628 (1981); Phys.\
Rev.\ B {\bf 26}, 5033 (1982).

\bibitem{evertz93}
H.~Evertz, G.~Lana, and M.~Marcu, Phys.\ Rev.\ Lett.\ {\bf 70}, 875 (1993).

\bibitem{kandel91} For a review of various methods, see
D.~Kandel and E.~Domany, Phys.\ Rev.\ B {\bf 43}, 8539 (1991).

\bibitem{hirsch83} J.E.~Hirsch and D.J.~Scalapino, Phys.\ Rev.\ Lett.\
{\bf 50}, 1168 (1983); Phys.\ Rev.\ B {\bf 29}, 5554 (1984).

\bibitem{suzuki76} M.~Suzuki, Prog.\ Theor.\ Phys.\ {\bf 56}, 1454 (1976).

\bibitem{blankenbecler81} R.~Blankenbecler, R.L.~Sugar, and D.J.~Scalapino,
Phys.\ Rev.\ D {\bf 24}, 2278 (1981).

\bibitem{okabe86} Y.~Okabe and M.~Kikuchi, Phys.\ Rev.\ B {\bf 34},
7896 (1986).

\bibitem{marcu87} M.~Marcu, J.~M\" uller, and F.-K.~Schimatzer,
Computer Phys.\ Commun.\ {\bf 44}, 63 (1987).

\bibitem{somsky92} W.R.~Somsky and J.E.~Gubernatis, Computers in
Physics {\bf 6}, 178 (1992).

\bibitem{binder88} K.~Binder and D.W.~Heermann, {\it Monte Carlo
Simulation in Statistical Physics} (Springer-Verlag, Heidelberg,
1988), Chap.~2.

\bibitem{allen87} M.P.~Allen and D.J.~Tildesley, {\it Computer
Simulations of Liquids} (Oxford University Press, Oxford, 1987),
Chap.~6.

\bibitem{hammersley64} J.M.~Hammersley and D.C.~Handcomb, {\it Monte
Carlo Methods} (Methuen, London, 1964), Chap.~5.

\bibitem{baxter82} R.J.~Baxter, {\it Exactly Solved Models in
Statistical Mechanics} (Academic Press, London, 1982), Chap.~8.

\bibitem{evertz} H.G.~Evertz and M.~Marcu. in {\it Computer
Simulations in Condensed Matter Physics VI}, edited by D.P.~ Landau,
K.K.~Mon, and H.-B.~Sch\" utterr (Springer-Verlag, Berlin, 1993), p.\
109.

\bibitem{kawashima94a} N.~Kawashima and J.E.~Gubernatis, ``Dual
Monte Carlo and cluster algorithms,'' preprint.

\bibitem{cases}   These exceptional case are: 1M MCS for four
simulations with algorithm $P$ at quarter-filling when $N=32$ and
$U=4$; 8M MCS for a simulation also with algorithm $P$ now at
half-filling but again when $N=32$ and $U=4$; and 10M MCS for
simulation with algorithm a canonical version of $PL$ at half-filling
when $N=32$ and $U=8$.

\bibitem{loh90} E.Y.~Loh,~Jr., J.E.~Gubernatis, R.T.~Scaletarr,
S.R.~White, D.J.~Scalapino, and R.L.~Sugar, Phys.\ Rev.\ B {\bf 41},
9301 (1990).

\bibitem{wiese93a} U.-J.~Wiese, ``Bosonization and cluster updating of
lattice fermions,'' preprint.

\bibitem{wiese93b} U.-J.~Wiese and H.-P.~Ying, ``A determination of
the low energy parameters of the 2-d Heisenberg antiferromagnet,''
preprint.

\bibitem{kawashima94b} N.~Kawashima and J.E.~Gubernatis, ``Cluster
algorithms for quantum spin simulations,'' in preparation.

\bibitem{scalapino} D.J.~Scalapino, S.R.~White, and S.C.~Zhang, Phys.\
Rev.\ B {\bf 47}, 7995 (1993) and references therein.

\end{references}
\end{document}